\documentclass[iop]{emulateapj}
\usepackage{apjfonts}

\usepackage[tight]{subfigure}
 \usepackage{amssymb, amsmath}
\usepackage{graphicx}
\usepackage{wasysym}
\usepackage{color}
\usepackage{verbatim}
\usepackage{hyperref}
\usepackage{xcolor}
\hypersetup{
    colorlinks,
    linkcolor={red!50!red},
    citecolor={blue!50!blue},
    urlcolor={blue!80!black}
}

\newcommand{\areport}{{\sc arepo-rt}}
\newcommand{\arepo}{{\sc arepo}}

\DeclareOldFontCommand{\rm}{\normalfont\rmfamily}{\mathrm}
\newcommand {\hi} {\ifmmode \ion{H}{I} \else $\ion{H}{I}$ \fi}
\newcommand {\hii} {\ifmmode \ion{H}{II} \else $\ion{H}{II}$ \fi}
\newcommand {\hei} {\ifmmode \ion{He}{I} \else $\ion{He}{I}$ \fi}
\newcommand {\heii} {\ifmmode \ion{He}{II} \else $\ion{He}{II}$ \fi}
\newcommand {\heiii} {\ifmmode \ion{He}{III} \else $\ion{He}{III}$ \fi}
\newcommand {\HM} {\ifmmode \ion{H}{$_2$} \else $\ion{H}{$_2$}$ \fi}
\newcommand {\xh} {\ifmmode X_\text{H} \else $X_\text{H}$ \fi}
\newcommand {\nh} {\ifmmode n_\text{H} \else $n_\text{H}$ \fi}

\shorttitle{dusty galactic winds}
\shortauthors{Kannan et al.}

\begin{document}

\title{Dust entrainment in galactic winds}

\author{Rahul Kannan\altaffilmark{1}, Mark Vogelsberger\altaffilmark{2}, Federico Marinacci\altaffilmark{3}, Laura V. Sales\altaffilmark{4}, Paul Torrey\altaffilmark{5}, and Lars Hernquist\altaffilmark{1}}
\email{rahul.kannan@cfa.harvard.edu}

\altaffiltext{1}{Center for Astrophysics $|$ Harvard $\&$ Smithsonian, 60 Garden Street, Cambridge, MA 02138, USA}
\altaffiltext{2}{Department of Physics, Kavli Institute for Astrophysics $\&$ Space Research, MIT, Cambridge 02139, MA, USA} 
\altaffiltext{3}{Department of Physics $\&$ Astronomy, University of Bologna, via Gobetti 93/2, 40129 Bologna, Italy}
\altaffiltext{4}{Department of Physics $\&$ Astronomy, University of California, Riverside, 900 University Avenue, Riverside, CA 92521, USA}
\altaffiltext{5}{Department of Astronomy, University of Florida, 211 Bryant Space Sciences Center, Gainesville, FL 32611 USA}

\begin{abstract}

Winds driven by stellar feedback are an essential part of the galactic ecosystem and are the main mechanism through which low-mass galaxies regulate their star formation. These winds are generally observed to be multi-phase with detections of entrained neutral and molecular gas. They are also thought to enrich the circum-galactic medium around galaxies with metals and dust. This ejected dust encodes information about the integrated star formation and outflow history of the galaxy. 
It is therefore, important to understand how much dust is entrained and driven out of the disc by galactic winds. Here we demonstrate that stellar feedback is efficient in driving dust-enriched winds and eject enough material to account for the amount of extraplanar dust observed in nearby galaxies. The amount of dust in the wind depends on the sites from where they are launched, with dustier galaxies launching more dust enriched outflows. Moreover, the outflowing cold-dense gas is significantly more dust-enriched than the volume filling hot tenuous material, naturally reproducing the complex multiphase structure of the outflowing wind observed in nearby galaxies. These results provide an important new insight into the dynamics, structure, and composition of galactic winds and their role in determining the dust content of the extragalactic gas in galaxies.

%A significant amount of cosmic dust has been observed to reside outside of galaxies in the so-called circumgalactic medium (CGM). 
%This dust is mostly formed in cool evolved stars and supernova remnants, which reside well within the main body of galaxies. 
%Exactly how the dust gets transported out into the CGM and how long it remains there, are currently open questions. Here we demonstrate that stellar feedback is efficient in driving dust-enriched outflows that eject enough material to completely account for the amount of extragalactic dust observed in nearby galaxies. The amount of dust in the outflow depends on the sites from where they are launched, with dustier galaxies launching more dust-enriched outflows. Moreover, the outflowing neutral gas is significantly more dust-enriched than the volume filling ionized gas, naturally reproducing the complex multiphase structure of gaseous outflows observed in nearby galaxies. These findings provide novel theoretical insight into the dynamics, structure, and composition of galactic outflows and their role in determining the dust content of the CGM of galaxies.
\end{abstract}


\section{Introduction}
\label{sec:intro}

Extraplanar dust is typically observed in emission in the far infrared (FIR) and  sub-mm frequencies, which are sensitive to cool dust \citep{Hughes1990, Radovich2001, Roussel2010, Melendez2015, McCormick2018}. It is also observed in the form of highly structured absorbing clouds against the background stellar light \citep{Howk2000}. Starlight is also scattered by the dust, forming reflection nebulae (RN) which are detected in the ultraviolet (UV) bands \citep{Hoopes2005, HK2014, Seon2014, Hodges2016}. These observations have shown that the  dust and warm ionized gas occupy separate regions of space, representing distinct phases of the multiphase extraplanar gas. The volume filling factor of the material traced by extraplanar dust is also much smaller than that of the ionized gas \citep{Howk2000, Rossa2004}. Moreover, in the starbursting galaxy M82, the outflow velocity of dust is substantially lower than that of both ionized and molecular gas \citep{Yoshida2011}, indicating that dust grains in the wind are kinematically decoupled from the gas. These observations indicate that the dynamics of dusty outflows is complex. Galactic scale winds in low-mass haloes ($\mathrm{M}_\mathrm{halo} \lesssim 10^{12} \mathrm{M}_\odot$) are mainly launched by the energy injection of supernova (SN) explosions in the inter-stellar medium \citep[ISM;][]{Chevalier1985, Strickland2000}. These winds are capable of entraining cold molecular and neutral gas \citep{Heckman2017, Naab2017} and might also be able to sweep up dust and expel them into the CGM of galaxies. Dust grains might also reform by accretion of metals after the shocked outflowing wind has cooled below the dust sputtering temperature \citep{Richings2018, Richings2018b}. Radiation pressure by the continuum absorption and scattering of photons on dust grains is another mechanism that has been invoked to explain large-scale galactic winds in highly luminous starbursting galaxies and in high luminosity quasars \citep{Murray2005, Barnes2018}.

Modeling the physics and dynamics of these dusty winds is paramount to understand the complex interplay of star formation, feedback, gas outflows and inflows, that regulate the formation and evolution of galaxies. There has been a lot of effort into modeling galactic winds. Most current galaxy formation models are quite empirical with the efficiency and dynamics of the winds designed to reproduce the large-scale properties of galaxies \citep{Vogelsberger2014, Schaye2015}. These models launch hydrodynamically decoupled winds that lack the ability to model low temperature gas or dust implying that they are unable to accurately capture the structure of the wind. While novel models have been implemented in these simulations, to track the dust distribution in the ISM \citep{McKinnon2017, Li2019}, the artificial wind launching mechanisms imply that the mass loading of and the dust entrainment properties in the wind are set by the input parameters and are therefore, not independent predictions of the simulation. Recently, \citet{Aoyama2018, Aoyama2020} used a sophisticated dust model that can follow the grain size distribution of dust to model the extinction curves of Milky-Way like galaxies. However, they did not address how the dust is entrained in galactic winds. 

\begin{table*}
\centering
\begin{tabular}{ccccccccccccc}

\hline
 Galaxy & $M_{\text{halo}}$ & $v_{200}$ & c & $M_{\text{bulge}}$ & $M_{\text{disk}}$ & $r_d$ & $h$ &  $r_g$ & $f_\text{gas}$ & $Z_\mathrm{init}$ & $L_\mathrm{box}$\\[0.14cm]

   & [M$_\odot$] & [km s$^{-1}$] & & [M$_\odot$] & [M$_\odot$] & [kpc] & [pc] &  [kpc] &  & [$Z_\odot$]  & [kpc]\\
\hline 
  MW & $1.53 \times 10^{12}$ & $169$ & 12 &  $1.5 \times 10^{10}$ & $4.74 \times 10^{10}$ & $3.0$ & $300$ & $6.0$ & 0.16 & 1.0 & 600\\ [0.14cm]
   LMC & $1.09 \times 10^{11}$ & $70$ & 14 &  $1.8 \times 10^{8}$ & $2.3 \times 10^{9}$ & $1.4$ & $140$ & $2.8$ & 0.19 & 0.5 & 200\\ [0.14cm]
  %DW & $2.18 \times 10^{10}$ & $41$ & 15 &  $1.14 \times 10^{7}$ & $1.59 \times 10^{8}$ & $0.7$ & $140$ & $0.7$ & 0.86 & 0.2 \\ [0.14cm]
  \hline

\end{tabular}
\caption{The structural parameters of the MW and LMC galaxies.}
\label{tab:ics}
\end{table*}

\begin{figure*}
\includegraphics[width=0.99\textwidth]{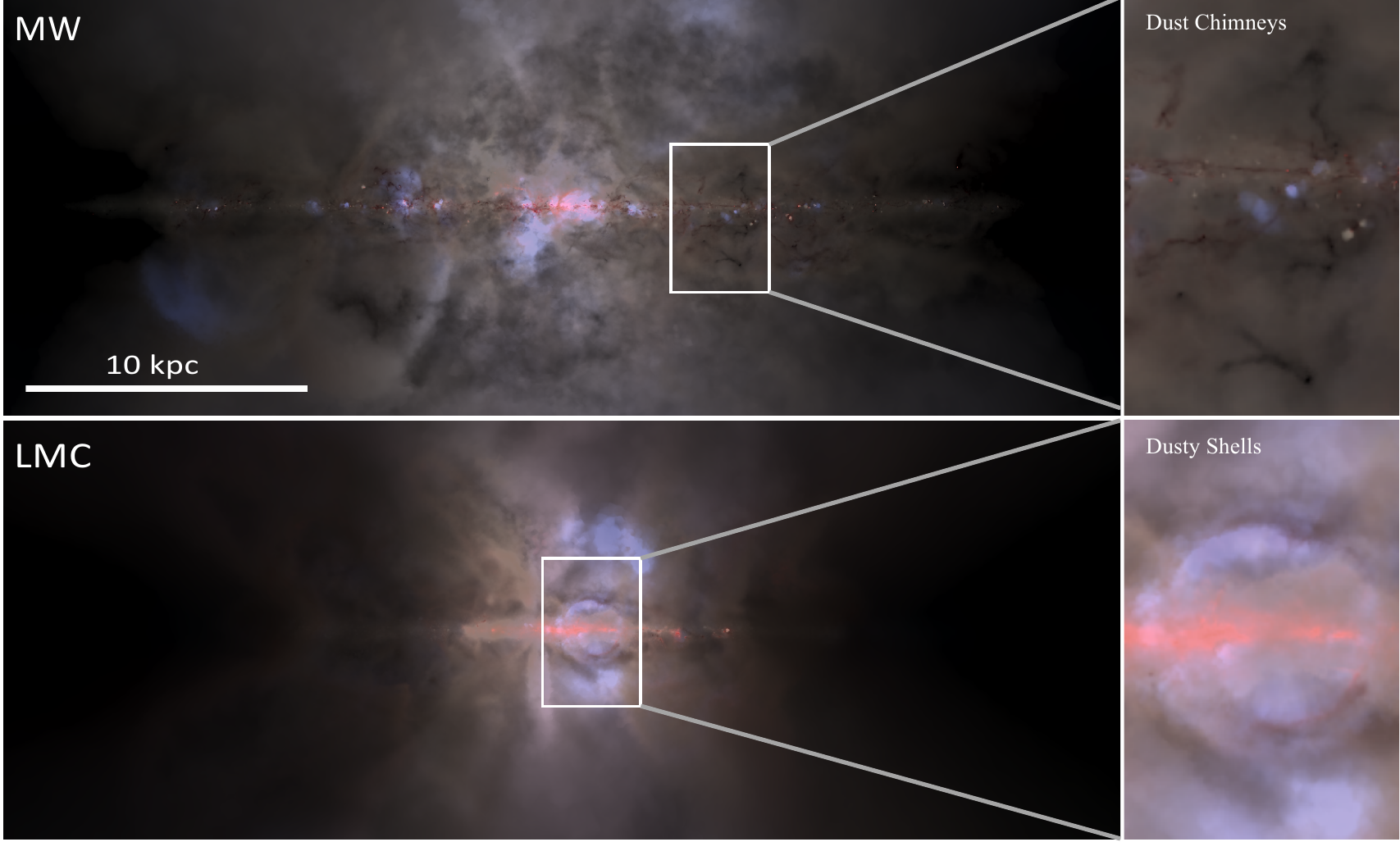}
\caption{A visualisation generated from the infrared (IR; red component), optical (yellow component) and the ionizing radiation (blue component) emission from the disk in the MW (top panel) and LMC (bottom panel) simulations as seen edge on. Dusty outflows and expanding dusty shells and chimneys are visible in both simulations. The insets show magnified examples of these topological features. These features are a clear sign of stellar feedback driven dusty outflows in galaxies \cite{Howk2000}.}
\label{fig:rgb}
\end{figure*}

In this Letter, we present high-resolution simulations of  isolated galaxies that self-consistently account for dust production and destruction mechanisms such as enrichment by SN and AGB stars, metal deposition onto dust grains, destruction in supernova remnants, thermal sputtering and dust-radiation coupling. We investigate the ability of stellar feedback driven winds in transporting dust from the ISM to the outer parts of the galactic halo.  Our methodology is introduced in
Section~\ref{sec:methods}, the main results are presented in
Section~\ref{sec:results} and finally, our conclusions are given
in Section~\ref{sec:conc}.

\section{Simulations}
\label{sec:methods}
The simulations are performed with \areport \, \citep{Kannan2019} a novel radiation hydrodynamic (RHD) extension of the moving mesh hydrodynamic code \arepo \, \citep{Springel2010}. The reduced speed of light approximation with ${\tilde c} = 10^3 \ \mathrm{km \ s}^{-1}$ is used in order to reduce the computational time.

\begin{figure*}
\includegraphics[width=0.99\textwidth]{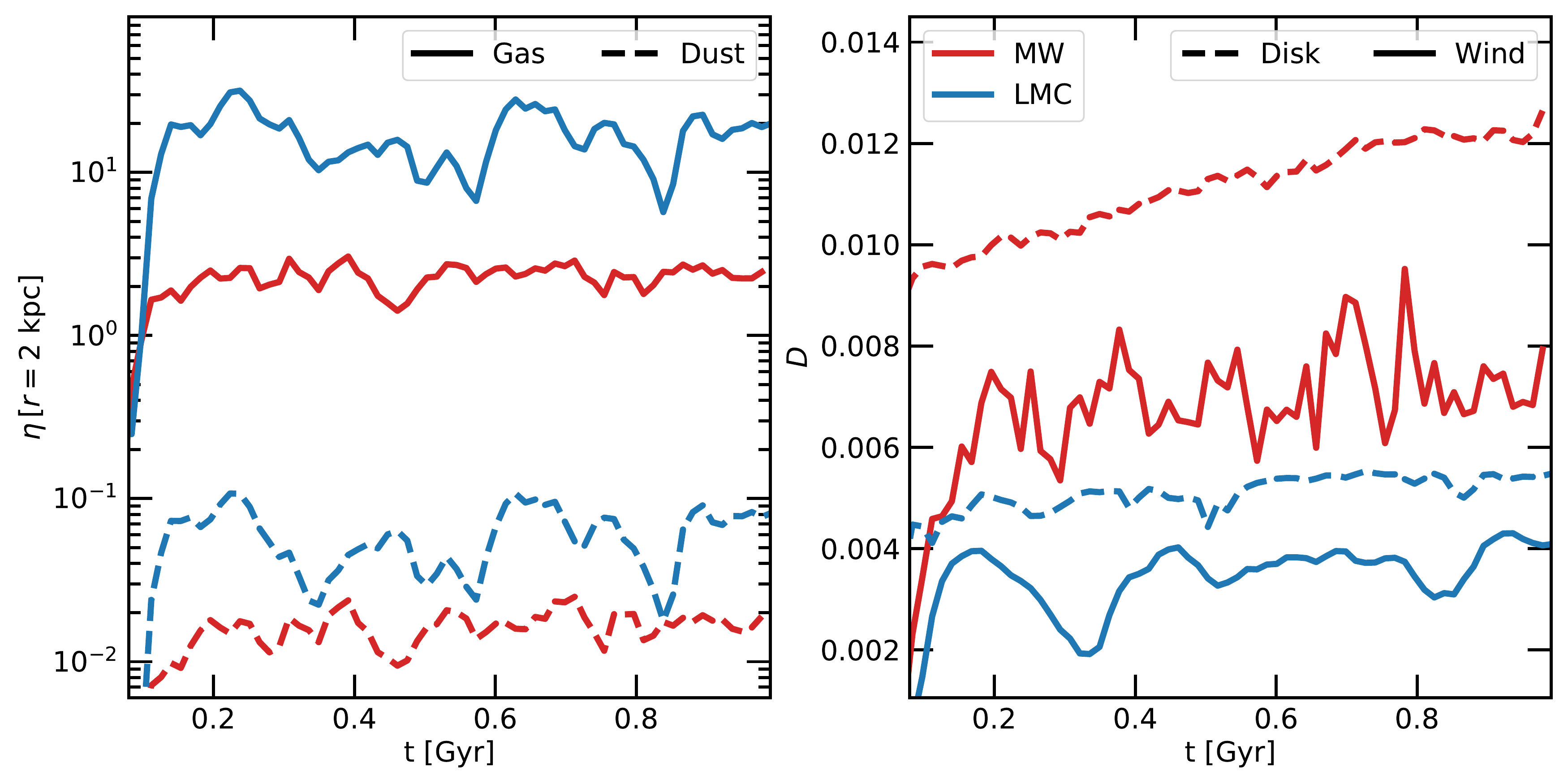}
\caption{The gas mass (solid curves) and dust mass (dashed curves) loading factors (left panel), and the dust-to-gas ratio ($D$) of the outflow (solid curves) and the disk (dashed curves) as a function of simulation time (right panel) in the MW (red curves) and LMC simulations (blue curves).  The gas mass loading factor hovers at $\sim 2$ for the MW simulation (red curves) and at $\sim 15$ for the LMC simulation (blue curves) throughout the duration of the simulation in agreement with previous rough theoretical estimates \cite{Muratov2015}. On the other hand, the mass loading factor of the dust predicted by the model is lower by about a factor of $200-500$. The plot also shows that the dustier galactic disk is able to launch more dust enriched outflows.}
\label{fig:mload}
\end{figure*} 

Gas cooling is implemented according the model described in \citet{Kannan2019b}. The simulations employ a novel self-consistent dust formation and destruction model \citep{McKinnon2017}. It accounts for three distinct dust production channels namely, SNII, SNIa and Asymptotic giant branch (AGB) stars \citep{Dwek1998}. The dust is assumed to be dynamically coupled to the gas and is passively advected along with the gas. The mass of dust in the ISM increases due to the gas-phase elements colliding with existing grains \citep{Dwek1998} and decrease due to shocks from SN remnants \citep{McKee1989} and sputtering in high temperature gas \citep{Tsai1995}.  

Star formation and feedback closely follow the implementation of \citet{Kannan2019b}  and \citet{Marinacci2019}. Cold gas above a density threshold $n_\text{th} = 10^3 \, \mathrm{cm}^{-3}$, is converted in star particles using the usual probabilistic approach. Additionally, we impose the condition that the star forming gas cloud needs to be self gravitating in order to form stars and add a Jeans pressure floor in order to stop under-resolved regions from fragmenting artificially. Three feedback mechanisms related to young stars, radiative feedback, stellar winds from young O, B and AGB stars and finally SN feedback are implemented. Photoheating, radiation pressure and photoelectric heating are modeled self-consistently through the radiative transfer scheme. Stellar winds from two classes of stars, massive, short-lived OB ($\lesssim 8 \ \text{M}_\odot$) stars and asymptotic giant branch (AGB) stars are included. SN is modelled using a boosted momentum injection method that compensates for the cooling loses that occur due to the inability to resolve the Sedov-Taylor phase.  The discrete nature of SN explosions is also modelled by imposing a time-step constraint for each stellar particle based on its age (i.e. evolutionary stage), such that the expectation value for the number of SN events per timestep is of the order of unity.

We run two high resolution isolated simulations of a Milky-Way like galaxy (MW; $M_\mathrm{halo} = 1.53 \times 10^{12} \, M_\odot$) and a Large Magellanic Cloud-like galaxy  (LMC; $M_\mathrm{halo} = 1.09 \times 10^{11} \, M_\odot$). We set up an equilibrium galaxy model consisting of a dark matter halo, a bulge and a stellar and gaseous disks in a domain of size $L_{\rm box}$. The DM halo and the bulge are modeled with a Hernquist profile \citep{Hernquist1990, Springel2005}. The gas and the stellar disk have an exponential profile in the radial direction with an effective radius $r_g$ and $r_d$ respectively. The vertical profile of the stellar disk follows a ${\rm sech}^2$ functional form with the scale height $h$. The vertical profile of the gaseous disk is computed self-consistently to ensure hydrostatic equilibrium at the beginning of the simulation. The initial gas temperature is set to $10^4$~K. The gas in the disk has a metallicity equal to a value of $Z_\mathrm{init}$. The lack of cosmological gas inflow into the disk can generate unrealistic gas metallicities. To avoid this the material returned from stars to the ISM has the same chemical composition of the star particle (i.e. production of new heavy elements is turned off). In this way the initial metallicity $Z_\mathrm{init}$ does not increase with time. The dark matter halo is modelled as a static background gravitational field, that is not impacted by the baryonic physics. The structural parameters of the galaxies under consideration are given in Table~\ref{tab:ics}. Each galaxy setup is run with a stellar mass resolution of $2.8 \times 10^3 \, M_\odot$ and gas mass resolution of $1.4 \times 10^3 \, M_\odot$. The corresponding gravitational softening lengths are $\epsilon_\star = 7.1 \, \mathrm{pc}$ and $\epsilon_\mathrm{gas} = 3.6 \, \mathrm{pc}$ respectively.

\section{Results}
\label{sec:results}

\begin{figure*}
\includegraphics[width=0.99\textwidth]{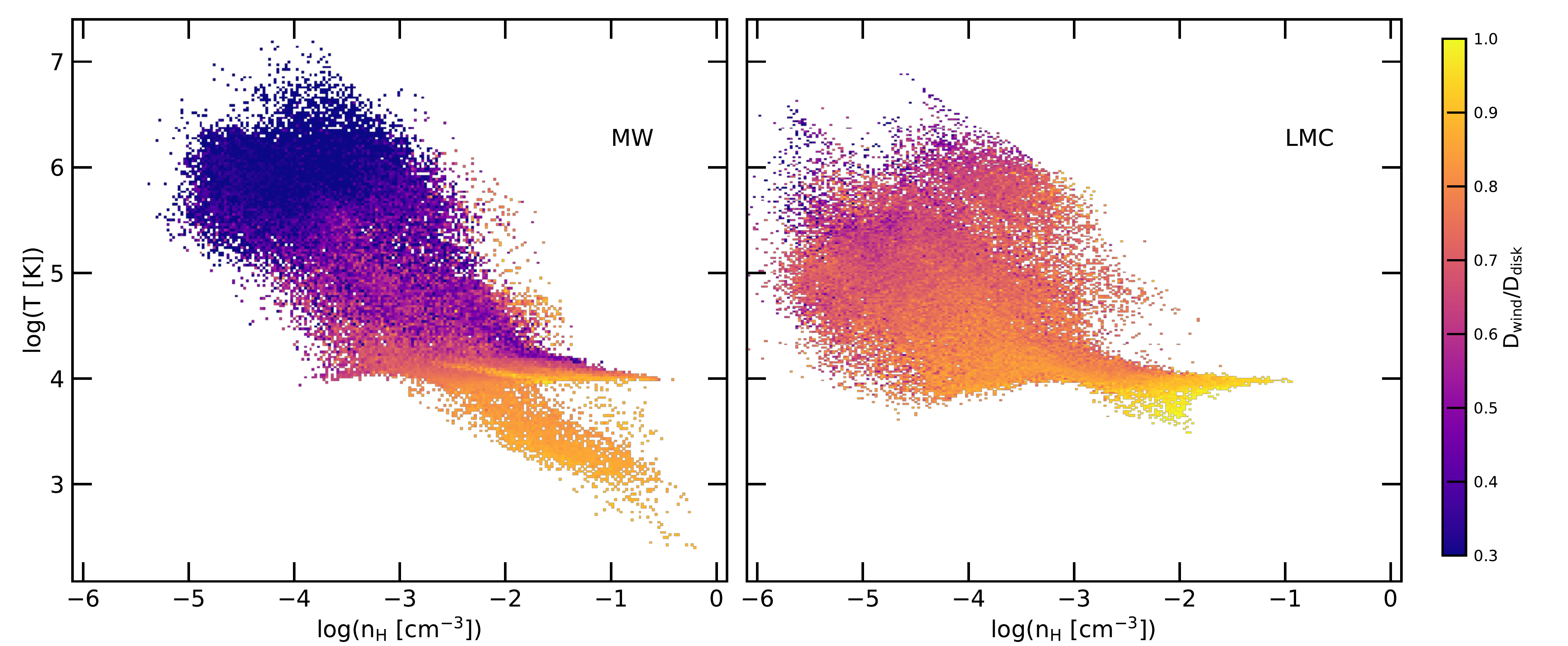}
\caption{A 2D histogram showing the relative dust-to-gas ratio (D$_{\rm outflow}$/D$_{\rm disk}$)  of the wind material ($2 \leq z \ (\mathrm{kpc}) < 10$) as a function of the density and temperature for the MW (left panel) and LMC (right panel) simulations.}
\label{fig:dgrout}
\end{figure*}

Fig.~\ref{fig:rgb} presents a visualization generated from the infrared (IR; red component), optical (yellow component) and the ionizing radiation (blue component) emission from the disk in the MW (top panel) and LMC (bottom panel) simulations as seen edge on. These visualizations demonstrate the manifold dusty extraplanar gas, extending above and below the disk. A variety of complex dust topologies can be observed including expanding, IR bright, dusty shells, around UV bright regions and a rich complex of filaments and chimney-like features (see magnified insets) that extends up to $\sim 2-3$~kpc above and below the plane of the galaxy. Such features have also been observed in nearby galaxies and have been attributed to hydrodynamical outflows driven by star formation activity in the disk \citep{Howk1999, Howk2000}.

%\begin{figure}
%\includegraphics[width=0.49\textwidth]{Figure/dgrout.pdf}
%\caption{The relative dust-to-gas ratio (top panels), outflow dust density (middle panels) and velocity of the gas and dust (bottom panels) split into neutral (purple curves) and ionized (brown curves) phases of the outflow, for the MW (left panels) and LMC (right panels) simulations. The neutral outflow is more dust enriched than the ionized outflowing material, with the density of dust in the neutral medium about $1-3$ orders of magnitude larger than the dust in the ionized gas.}
%\label{fig:dgrout}
%\end{figure}

 A quantitative picture of wind is obtained by measuring the gas mass (solid curves) and dust mass loading (dashed curves) factors ($\eta$), defined as the ratio of the mass outflow rate to the star formation rate of the galaxy, as a function of simulation time (see left panel of Fig.~\ref{fig:mload}). These quantities are measured at a height of $2$~kpc from the plane of the disk. The gas mass loading factor hovers at $\sim 2$ for the MW (red curves) and at $\sim 15$ for the LMC run (blue curves) throughout the duration of the simulation. This is in agreement with previous rough theoretical estimates \citep{Muratov2015}. On the other hand, the loading factor of the dust predicted by the model is lower by about a factor of $200-500$. These low dust loading factors are mainly due to the fact that the dust-to-gas ratio of the disk, from where these outflows are launched, is usually less than $1\%$. This is demonstrated more clearly in the right panel of Fig.~\ref{fig:mload}, which shows the dust-to-gas ratio ($D$) of the wind (calculated as the mass loading of dust divided by the mass loading of gas; solid curves) compared to the dust-to-gas ratio of the central disk (dashed curves). $D$ of the MW disk is about a factor of two higher than the disk in the LMC simulation. This is because the metallicity of the gas in the MW simulation is initialised to the canonical solar abundance value ($Z_\odot$) while the metallicity of the LMC run is $Z_\odot / 2$, in order to match the observational estimates \citep{Madden2013}. Although the simulations are initialised without dust, the gas of the disk is enriched during the course of the simulation, such that the average $D$ is $\sim 0.01$ and $\sim 4 \times 10^{-3}$ for the MW and LMC disks, respectively, which is in excellent agreement with observational estimates \citep{Duval2014, Giannetti2017}. The figure shows that the exact level of dust enrichment in outflows depends on the dust abundance in the disk from which the winds are launched, with the dustier disk launching more enriched winds. The fraction of dust in the disk is always higher than in the outflow, by about $30-40\%$, meaning that the outflows are less dust enriched than the sites from which they are launched. This is due to the fact that a certain fraction of the dust is destroyed in SN shocks and thermal sputtering in the high temperature gas resulting from SN explosions. We note that turning off the radiation pressure does not change the mass loading factors of either the dust or the gas in both the MW and LMC simulations. This is because the star formation rates of these galaxies are moderately low (MW: $\sim 2 \, \mathrm{M}_\odot \, \mathrm{yr}^{-1}$; LMC: $\sim 0.1 \, \mathrm{M}_\odot \, \mathrm{yr}^{-1}$), such that the light-to-mass ratio is not large enough to launch winds via radiation pressure. This is consistent with results of previous theoretical consideration \citep{Rosdahl2015, Kannan2019c} that showed that radiation pressure is only  important in dense, compact, and rapidly star-forming galaxies like the M82 \citep{Thompson2015}. These results demonstrate that dust is efficiently entrained and expelled out of the disk by supernova driven outflows, offering a path to explain the transport of dust from galaxies into the circum-galactic medium.
%making them  potentially an important process by which the the CGM is enriched with dust.

 A thorough understanding of the gas and dust structure in the wind is obtained by plotting a two-dimensional histogram of the relative dust to gas ratio (D$_{\rm wind}$/D$_{\rm disk}$) as a function of the gas temperature and density (Fig.~\ref{fig:dgrout}) in the MW (left panel) and LMC (right panel) simulations. We consider all outflowing gas that is at a height between two and ten kpc from the midplane of the disk. Interestingly, both the MW and LMC simulations show a complex wind structure that contains both a cold dense ($100 \lesssim {\rm T [K]} \lesssim 10^4$; $10^{-3} \lesssim {\rm n_H [cm}^{-3}] \lesssim 1.0$) and a hot tenuous phase that do not necessarily co-exist in pressure equilibrium. $D$ shows a very strong dependence on the temperature of the gas. On the other hand, almost all the density dependence of $D$ arises from the fact that the cold gas needs to be necessarily dense in order to self-shield from the local radiation background. The $D$ of the cold-dense gas in the wind is only slightly lower than $D$ of the disk it is launched from ($D_\mathrm{outflow}/D_\mathrm{disk} \sim 0.8 -1.0$), while $D$ of the warm low-density material is generally lower by about a factor of $2-3$ ($D_\mathrm{outflow}/D_\mathrm{disk} \sim 0.3-0.5$). Warm gas is primarily ejected from the sites of SN explosions, which are efficient at destroying dust due to the shocks from SN remnants \citep{McKee1989}. Therefore, $D$ of the hot low density outflow is much lower than $D$ of the disk. The dense material on the other hand, arises from the hot wind impinging on cold clouds and accelerating them to form a multiphase, kinematically decoupled wind \citep{Klein1994, Schwartz2004}. This cold and dense entrained material, unaffected by shocks, is more suitable for dust grains to survive, allowing for it to be almost as dust enriched as the disk it has been launched from. We are, therefore, able to reproduce a realistic wind, that is multiphase, with the dense outflowing material being more dust enriched than the volume filling warm wind \citep{Rossa2004, HK2016}.

\begin{figure}
\begin{center}
\includegraphics[width=0.49\textwidth]{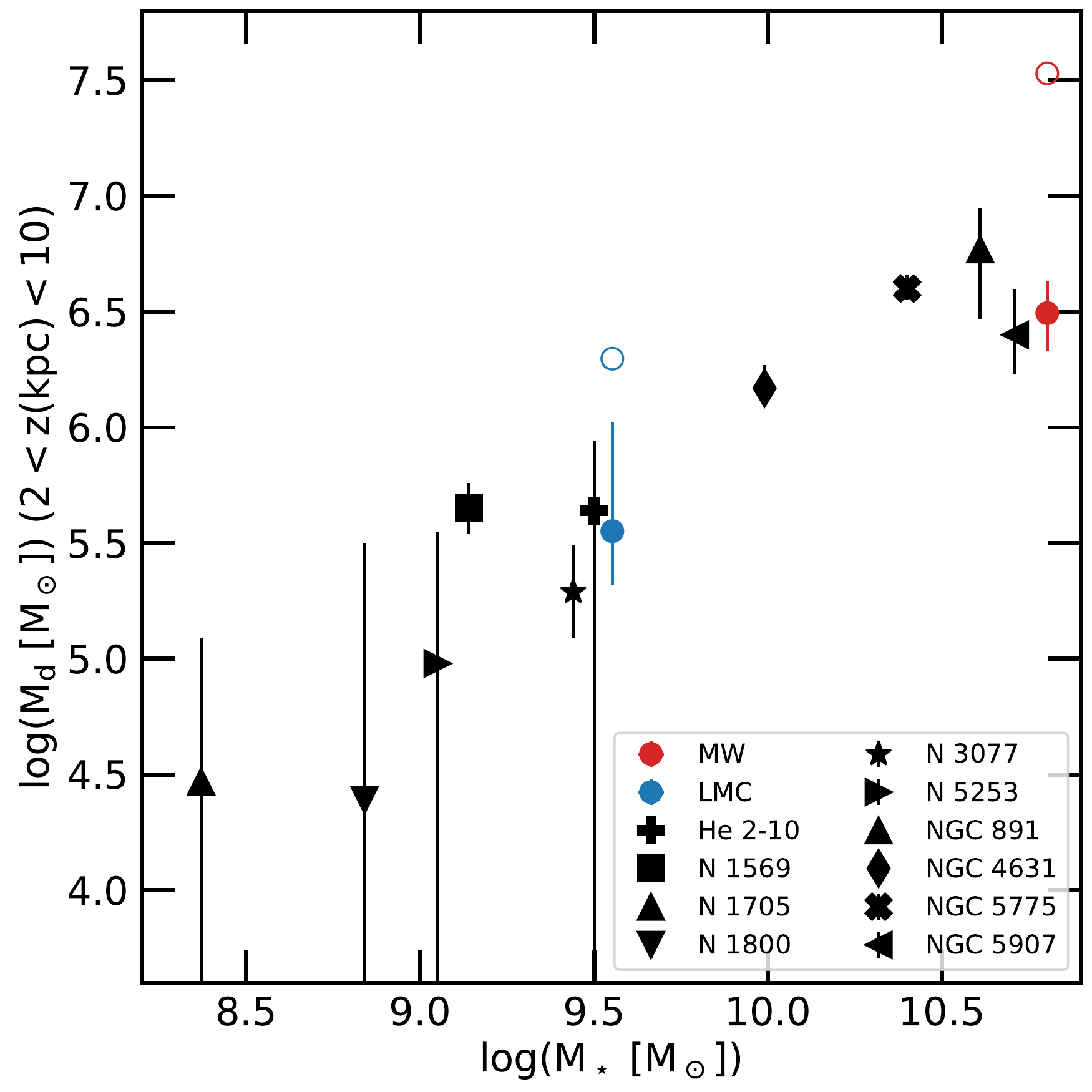}
\caption{The total amount of extraplanar dust mass ($2 \leq z \ (\mathrm{kpc}) < 10$) as a function of the total stellar mass of the galaxy in the MW (red point) and LMC  (blue point) simulations. For comparison we also show the observational estimates from nearby dwarf \citep[][N1705, N1800, N5253, N1569, He2-10, N3077]{McCormick2018} and spiral \citep[][NGC 891, NGC 4631, NGC 5775, NGC 5907]{HK2016}galaxies. The corresponding unfilled circles show the total amount of dust mass ejected from the disk during the entire duration of the simulation time ($1$~Gyr), which is about an order of magnitude larger than the current amount of extraplanar dust in the galaxy.}
\label{fig:cgmdust}
\end{center}
\end{figure}

  Finally, Figure~\ref{fig:cgmdust}, shows the amount of dust mass ejected out of the disk of the galaxy ($2 \leq z \ (\mathrm{kpc}) < 10$)  by the galactic wind, as a function of the stellar mass of the galaxy in the MW (filled red circle) and LMC  (filled blue circle) simulations compared to the observational estimates for the nearby dwarf \citep{McCormick2018} and spiral galaxies \citep{HK2016}.   The error bars show the time variation in the amount of dust after the disk in the simulation has settled down ($t>0.2$~Gyr). Despite the fact that dust masses of the dwarf (emission in the far-IR from {\it Herschel}) and spiral (reflected UV light of dust grains with {\it GALEX} and {\it Swift}) galaxies are obtained using very different techniques, they do show a relatively tight correlation with the stellar mass of the galaxy.
  
  The amount of extraplanar dust in the MW and LMC galaxies are about $\sim 3 \times 10^6 \, M_\odot$ and $\sim 3 \times 10^5 \, M_\odot$ respectively, which is in agreement with the observational estimates. We note that the total amount of dust mass ejected (above $2$~kpc but not above $10$~kpc) from the disk of the galaxy integrated over the entire duration of the simulation time ($1$~Gyr; unfilled circles) is about a factor of $\sim 10$ higher than current mass of the extraplanar dust found in the galaxies. This is because most of the dust ejected by the galaxies falls back down as part of the fountain flow and only a small fraction of it remains outside the disk of the galaxy at any given point in time. This implies that the amount of extraplanar dust depends chiefly on the current/recent star formation rate of the galaxy and the corresponding small scale fountain flow setup by it, and should therefore be largely independent of the cosmological environment of the galaxy. Higher mass galaxies generally have higher star formation rates, which leads to larger outflow rates. They also contain dustier discs, which launch more dust enriched outflows. These two factors combine to generate a positive stellar mass correlation with the extraplanar dust mass of the galaxy.  
\section{Conclusions}
\label{sec:conc}

We have presented high resolution isolated simulations of a Milky-Way like (MW) and a Large Magellanic Cloud like (LMC) galaxies  with detailed modeling of dynamic dust formation, evolution, and destruction. Using these simulations we investigate the level of dust entrainment in stellar feedback driven winds. Our main results are summarised as follows.
\begin{enumerate}
    \item The gas mass loading factor for the MW and LMC simulations is about $\sim 2$ and $\sim 15$ respectively, in agreement with previous estimates \citep{Muratov2015}. On the other hand, we predict that the dust mass loading factor should be lower by about 200-500. 
    \item The exact level of dust enrichment in the wind depends on the sites from where they are launched with the dustier gas disk launching more dust enriched outflows.
    \item The $D$ of the cold-dense gas in the wind is only slightly lower than $D$ of the disk it is launched from, while $D$ of the warm low-density material is generally lower by about a factor of $2-3$.  This is because, the warm gas is primarily ejected from the sites of SN explosions, which are efficient at destroying dust due to the shocks from SN remnants. The dense entrained material on the other hand is unaffected by shocks and is therefore more suitable for dust grains to survive.
    \item The galactic winds launch enough material to completely account for the amount of extraplanar dust observed in nearby dwarf and spiral galaxies.
    \item Most of the ejected dust participates in a small scale fountain flow and only a small fraction of it remains outside the plane of the disk at any given point.
\end{enumerate}

We can therefore conclude that stellar feedback driven outflows expel enough dust from the disc of the galaxies to explain the amount of extraplanar dust in galaxies observed in the nearby Universe. In the future we plan to extend this study to a fully cosmological framework which will allow us comprehensively quantify the dust content in the circum-galactic and intergalactic gas in the Universe.  These findings will be especially important for current and upcoming sub-mm, FIR and UV facilities like {\it Herschel}, {\it ALMA} and {\it JWST} that will help to detect dusty winds  \citep{HK2019} in a variety of galaxies at various redshifts.

\begin{acknowledgements}
FM is supported by the Program "Rita Levi Montalcini" of the Italian MIUR. LVS is thankful for the financial support from NASA through HST-AR-14582. Computing  resources  supporting  this  work  were  provided  by  the  NASA  High-End  Computing (HEC) Program through the NASA Advanced Supercomputing (NAS) Division at Ames Research Center.
\end{acknowledgements}

\bibliographystyle{apj} 
\bibliography{DustyWinds.bib}

\begin{thebibliography}{}
\expandafter\ifx\csname natexlab\endcsname\relax\def\natexlab#1{#1}\fi
\providecommand{\url}[1]{\href{#1}{#1}}

\bibitem[{{Aoyama} {et~al.}(2020){Aoyama}, {Hirashita}, \&
  {Nagamine}}]{Aoyama2020}
{Aoyama}, S., {Hirashita}, H., \& {Nagamine}, K. 2020, \mnras, 491, 3844

\bibitem[{{Aoyama} {et~al.}(2018){Aoyama}, {Hou}, {Hirashita}, {Nagamine}, \&
  {Shimizu}}]{Aoyama2018}
{Aoyama}, S., {Hou}, K.-C., {Hirashita}, H., {Nagamine}, K., \& {Shimizu}, I.
  2018, \mnras, 478, 4905

\bibitem[{{Barnes} {et~al.}(2018){Barnes}, {Kannan}, {Vogelsberger}, \&
  {Marinacci}}]{Barnes2018}
{Barnes}, D.~J., {Kannan}, R., {Vogelsberger}, M., \& {Marinacci}, F. 2018,
  arXiv e-prints, arXiv:1812.01611

\bibitem[{{Chevalier} \& {Clegg}(1985)}]{Chevalier1985}
{Chevalier}, R.~A., \& {Clegg}, A.~W. 1985, \nat, 317, 44

\bibitem[{{Dwek}(1998)}]{Dwek1998}
{Dwek}, E. 1998, \apj, 501, 643

\bibitem[{{Giannetti} {et~al.}(2017){Giannetti}, {Leurini}, {K{\"o}nig},
  {Urquhart}, {Pillai}, {Brand}, {Kauffmann}, {Wyrowski}, \&
  {Menten}}]{Giannetti2017}
{Giannetti}, A., {Leurini}, S., {K{\"o}nig}, C., {et~al.} 2017, \aap, 606, L12

\bibitem[{{Heckman} \& {Thompson}(2017)}]{Heckman2017}
{Heckman}, T.~M., \& {Thompson}, T.~A. 2017, arXiv e-prints, arXiv:1701.09062

\bibitem[{{Hernquist}(1990)}]{Hernquist1990}
{Hernquist}, L. 1990, \apj, 356, 359

\bibitem[{{Hodges-Kluck} \& {Bregman}(2014)}]{HK2014}
{Hodges-Kluck}, E., \& {Bregman}, J.~N. 2014, \apj, 789, 131

\bibitem[{{Hodges-Kluck} {et~al.}(2016{\natexlab{a}}){Hodges-Kluck},
  {Cafmeyer}, \& {Bregman}}]{Hodges2016}
{Hodges-Kluck}, E., {Cafmeyer}, J., \& {Bregman}, J.~N. 2016{\natexlab{a}},
  \apj, 833, 58

\bibitem[{{Hodges-Kluck} {et~al.}(2016{\natexlab{b}}){Hodges-Kluck},
  {Cafmeyer}, \& {Bregman}}]{HK2016}
---. 2016{\natexlab{b}}, \apj, 833, 58

\bibitem[{{Hodges-Kluck} {et~al.}(2019){Hodges-Kluck}, {Corrales}, {Veilleux},
  {Bregman}, {Li}, \& {Melendez}}]{HK2019}
{Hodges-Kluck}, E., {Corrales}, L., {Veilleux}, S., {et~al.} 2019, \baas, 51,
  249

\bibitem[{{Hoopes} {et~al.}(2005){Hoopes}, {Heckman}, {Strickland }, {Seibert},
  {Madore}, {Rich}, {Bianchi}, {Gil de Paz}, {Burgarella}, {Thilker},
  {Friedman}, {Barlow}, {Byun}, {Donas}, {Forster}, {Jelinsky}, {Lee},
  {Malina}, {Martin}, {Milliard}, {Morrissey}, {Neff}, {Schiminovich},
  {Siegmund}, {Small}, {Szalay}, {Welsh}, \& {Wyder}}]{Hoopes2005}
{Hoopes}, C.~G., {Heckman}, T.~M., {Strickland }, D.~K., {et~al.} 2005, \apjl,
  619, L99

\bibitem[{{Howk} \& {Savage}(1999)}]{Howk1999}
{Howk}, J.~C., \& {Savage}, B.~D. 1999, \aj, 117, 2077

\bibitem[{{Howk} \& {Savage}(2000)}]{Howk2000}
---. 2000, \aj, 119, 644

\bibitem[{{Hughes} {et~al.}(1990){Hughes}, {Gear}, \& {Robson}}]{Hughes1990}
{Hughes}, D.~H., {Gear}, W.~K., \& {Robson}, E.~I. 1990, \mnras, 244, 759

\bibitem[{{Kannan} {et~al.}(2020){Kannan}, {Marinacci}, {Simpson}, {Glover}, \&
  {Hernquist}}]{Kannan2019c}
{Kannan}, R., {Marinacci}, F., {Simpson}, C.~M., {Glover}, S. C.~O., \&
  {Hernquist}, L. 2020, \mnras, 491, 2088

\bibitem[{{Kannan} {et~al.}(2019{\natexlab{a}}){Kannan}, {Marinacci},
  {Vogelsberger}, {Sales}, {Torrey}, {Springel}, \& {Hernquist}}]{Kannan2019b}
{Kannan}, R., {Marinacci}, F., {Vogelsberger}, M., {et~al.} 2019{\natexlab{a}},
  arXiv e-prints, arXiv:1910.14041

\bibitem[{{Kannan} {et~al.}(2019{\natexlab{b}}){Kannan}, {Vogelsberger},
  {Marinacci}, {McKinnon}, {Pakmor}, \& {Springel}}]{Kannan2019}
{Kannan}, R., {Vogelsberger}, M., {Marinacci}, F., {et~al.} 2019{\natexlab{b}},
  \mnras, 485, 117

\bibitem[{{Klein} {et~al.}(1994){Klein}, {McKee}, \& {Colella}}]{Klein1994}
{Klein}, R.~I., {McKee}, C.~F., \& {Colella}, P. 1994, \apj, 420, 213

\bibitem[{{Li} {et~al.}(2019){Li}, {Narayanan}, \& {Dav{\'e}}}]{Li2019}
{Li}, Q., {Narayanan}, D., \& {Dav{\'e}}, R. 2019, \mnras, 490, 1425

\bibitem[{{Madden} {et~al.}(2013){Madden}, {R{\'e}my-Ruyer}, {Galametz},
  {Cormier}, {Lebouteiller}, {Galliano}, {Hony}, {Bendo}, {Smith}, {Pohlen},
  {Roussel}, {Sauvage}, {Wu}, {Sturm}, {Poglitsch}, {Contursi}, {Doublier},
  {Baes}, {Barlow}, {Boselli}, {Boquien}, {Carlson}, {Ciesla}, {Cooray},
  {Cortese}, {de Looze}, {Irwin}, {Isaak}, {Kamenetzky}, {Karczewski}, {Lu},
  {MacHattie}, {O'Halloran}, {Parkin}, {Rangwala}, {Schirm}, {Schulz},
  {Spinoglio}, {Vaccari}, {Wilson}, \& {Wozniak}}]{Madden2013}
{Madden}, S.~C., {R{\'e}my-Ruyer}, A., {Galametz}, M., {et~al.} 2013, \pasp,
  125, 600

\bibitem[{{Marinacci} {et~al.}(2019){Marinacci}, {Sales}, {Vogelsberger},
  {Torrey}, \& {Springel}}]{Marinacci2019}
{Marinacci}, F., {Sales}, L.~V., {Vogelsberger}, M., {Torrey}, P., \&
  {Springel}, V. 2019, arXiv e-prints, arXiv:1905.08806

\bibitem[{{McCormick} {et~al.}(2018){McCormick}, {Veilleux}, {Mel{\'e}ndez},
  {Martin}, {Bland -Hawthorn}, {Cecil}, {Heitsch}, {M{\"u}ller}, {Rupke}, \&
  {Engelbracht}}]{McCormick2018}
{McCormick}, A., {Veilleux}, S., {Mel{\'e}ndez}, M., {et~al.} 2018, \mnras,
  477, 699

\bibitem[{{McKee}(1989)}]{McKee1989}
{McKee}, C. 1989, in IAU Symposium, Vol. 135, Interstellar Dust, ed. L.~J.
  {Allamandola} \& A.~G.~G.~M. {Tielens}, 431

\bibitem[{{McKinnon} {et~al.}(2017){McKinnon}, {Torrey}, {Vogelsberger},
  {Hayward}, \& {Marinacci}}]{McKinnon2017}
{McKinnon}, R., {Torrey}, P., {Vogelsberger}, M., {Hayward}, C.~C., \&
  {Marinacci}, F. 2017, \mnras, 468, 1505

\bibitem[{{Mel{\'e}ndez} {et~al.}(2015){Mel{\'e}ndez}, {Veilleux}, {Martin},
  {Engelbracht}, {Bland-Hawthorn}, {Cecil}, {Heitsch}, {McCormick},
  {M{\"u}ller}, {Rupke}, \& {Teng}}]{Melendez2015}
{Mel{\'e}ndez}, M., {Veilleux}, S., {Martin}, C., {et~al.} 2015, \apj, 804, 46

\bibitem[{{Muratov} {et~al.}(2015){Muratov}, {Kere{\v{s}}},
  {Faucher-Gigu{\`e}re}, {Hopkins}, {Quataert}, \& {Murray}}]{Muratov2015}
{Muratov}, A.~L., {Kere{\v{s}}}, D., {Faucher-Gigu{\`e}re}, C.-A., {et~al.}
  2015, \mnras, 454, 2691

\bibitem[{{Murray} {et~al.}(2005){Murray}, {Quataert}, \&
  {Thompson}}]{Murray2005}
{Murray}, N., {Quataert}, E., \& {Thompson}, T.~A. 2005, \apj, 618, 569

\bibitem[{{Naab} \& {Ostriker}(2017)}]{Naab2017}
{Naab}, T., \& {Ostriker}, J.~P. 2017, \araa, 55, 59

\bibitem[{{Radovich} {et~al.}(2001){Radovich}, {Kahanp{\"a}{\"a}}, \&
  {Lemke}}]{Radovich2001}
{Radovich}, M., {Kahanp{\"a}{\"a}}, J., \& {Lemke}, D. 2001, \aap, 377, 73

\bibitem[{{Richings} \&
  {Faucher-Gigu{\`e}re}(2018{\natexlab{a}})}]{Richings2018}
{Richings}, A.~J., \& {Faucher-Gigu{\`e}re}, C.-A. 2018{\natexlab{a}}, \mnras,
  474, 3673

\bibitem[{{Richings} \&
  {Faucher-Gigu{\`e}re}(2018{\natexlab{b}})}]{Richings2018b}
---. 2018{\natexlab{b}}, \mnras, 478, 3100

\bibitem[{{Roman-Duval} {et~al.}(2014){Roman-Duval}, {Gordon}, {Meixner},
  {Bot}, {Bolatto}, {Hughes}, {Wong}, {Babler}, {Bernard}, {Clayton}, {Fukui},
  {Galametz}, {Galliano}, {Glover}, {Hony}, {Israel}, {Jameson},
  {Lebouteiller}, {Lee}, {Li}, {Madden}, {Misselt}, {Montiel}, {Okumura},
  {Onishi}, {Panuzzo}, {Reach}, {Remy-Ruyer}, {Robitaille}, {Rubio}, {Sauvage},
  {Seale}, {Sewilo}, {Staveley-Smith}, \& {Zhukovska}}]{Duval2014}
{Roman-Duval}, J., {Gordon}, K.~D., {Meixner}, M., {et~al.} 2014, \apj, 797, 86

\bibitem[{{Rosdahl} {et~al.}(2015){Rosdahl}, {Schaye}, {Teyssier}, \&
  {Agertz}}]{Rosdahl2015}
{Rosdahl}, J., {Schaye}, J., {Teyssier}, R., \& {Agertz}, O. 2015, \mnras, 451,
  34

\bibitem[{{Rossa} {et~al.}(2004){Rossa}, {Dettmar}, {Walterbos}, \&
  {Norman}}]{Rossa2004}
{Rossa}, J., {Dettmar}, R.-J., {Walterbos}, R. A.~M., \& {Norman}, C.~A. 2004,
  \aj, 128, 674

\bibitem[{{Roussel} {et~al.}(2010){Roussel}, {Wilson}, {Vigroux}, {Isaak},
  {Sauvage}, {Madden}, {Auld}, {Baes}, {Barlow}, {Bendo}, {Bock}, {Boselli},
  {Bradford}, {Buat}, {Castro-Rodriguez}, {Chanial}, {Charlot}, {Ciesla},
  {Clements}, {Cooray}, {Cormier}, {Cortese}, {Davies}, {Dwek}, {Eales},
  {Elbaz}, {Galametz}, {Galliano}, {Gear}, {Glenn}, {Gomez}, {Griffin}, {Hony},
  {Levenson}, {Lu}, {O'Halloran}, {Okumura}, {Oliver}, {Page}, {Panuzzo},
  {Papageorgiou}, {Parkin}, {Perez-Fournon}, {Pohlen}, {Rangwala}, {Rigby},
  {Rykala}, {Sacchi}, {Schulz}, {Schirm}, {Smith}, {Spinoglio}, {Stevens},
  {Srinivasan}, {Symeonidis}, {Trichas}, {Vaccari}, {Wozniak}, {Wright}, \&
  {Zeilinger}}]{Roussel2010}
{Roussel}, H., {Wilson}, C.~D., {Vigroux}, L., {et~al.} 2010, \aap, 518, L66

\bibitem[{{Schaye} {et~al.}(2015){Schaye}, {Crain}, {Bower}, {Furlong},
  {Schaller}, {Theuns}, {Dalla Vecchia}, {Frenk}, {McCarthy}, {Helly},
  {Jenkins}, {Rosas-Guevara}, {White}, {Baes}, {Booth}, {Camps}, {Navarro},
  {Qu}, {Rahmati}, {Sawala}, {Thomas}, \& {Trayford}}]{Schaye2015}
{Schaye}, J., {Crain}, R.~A., {Bower}, R.~G., {et~al.} 2015, \mnras, 446, 521

\bibitem[{{Schwartz} \& {Martin}(2004)}]{Schwartz2004}
{Schwartz}, C.~M., \& {Martin}, C.~L. 2004, \apj, 610, 201

\bibitem[{{Seon} {et~al.}(2014){Seon}, {Witt}, {Shinn}, \& {Kim}}]{Seon2014}
{Seon}, K.-i., {Witt}, A.~N., {Shinn}, J.-h., \& {Kim}, I.-j. 2014, \apjl, 785,
  L18

\bibitem[{{Springel}(2010)}]{Springel2010}
{Springel}, V. 2010, \mnras, 401, 791

\bibitem[{{Springel} {et~al.}(2005){Springel}, {Di Matteo}, \&
  {Hernquist}}]{Springel2005}
{Springel}, V., {Di Matteo}, T., \& {Hernquist}, L. 2005, \apjl, 620, L79

\bibitem[{{Strickland} {et~al.}(2000){Strickland}, {Heckman}, {Weaver}, \&
  {Dahlem}}]{Strickland2000}
{Strickland}, D.~K., {Heckman}, T.~M., {Weaver}, K.~A., \& {Dahlem}, M. 2000,
  \aj, 120, 2965

\bibitem[{{Thompson} {et~al.}(2015){Thompson}, {Fabian}, {Quataert}, \&
  {Murray}}]{Thompson2015}
{Thompson}, T.~A., {Fabian}, A.~C., {Quataert}, E., \& {Murray}, N. 2015,
  \mnras, 449, 147

\bibitem[{{Tsai} \& {Mathews}(1995)}]{Tsai1995}
{Tsai}, J.~C., \& {Mathews}, W.~G. 1995, \apj, 448, 84

\bibitem[{{Vogelsberger} {et~al.}(2014){Vogelsberger}, {Genel}, {Springel},
  {Torrey}, {Sijacki}, {Xu}, {Snyder}, {Bird}, {Nelson}, \&
  {Hernquist}}]{Vogelsberger2014}
{Vogelsberger}, M., {Genel}, S., {Springel}, V., {et~al.} 2014, \nat, 509, 177

\bibitem[{{Yoshida} {et~al.}(2011){Yoshida}, {Kawabata}, \&
  {Ohyama}}]{Yoshida2011}
{Yoshida}, M., {Kawabata}, K.~S., \& {Ohyama}, Y. 2011, \pasj, 63, 493

\end{thebibliography}

\end{document}